\documentclass[iop]{emulateapj}

\usepackage{natbib}
\usepackage{amsmath}

\newcommand{\myemail}{billylquarles@gmail.com}

\shorttitle{Stability in $\alpha$ Cen}
\shortauthors{Quarles, Lissauer, \& Kaib }

\begin{document}


\title{Long-Term Stability of Planets in the $\alpha$ Centauri System, II: Forced Eccentricities}


\author{B. Quarles}
\affil{HL Dodge Department of Physics \& Astronomy, University of Oklahoma, Norman, OK 73019,
USA}
\email{\myemail}
\author{Jack J. Lissauer}
\affil{NASA Ames Research Center, Space Science and Astrobiology Division MS 245-3,
    Moffett Field, CA 94035}
\author{N. Kaib}
\affil{HL Dodge Department of Physics \& Astronomy, University of Oklahoma, Norman, OK 73019,
USA}


\begin{abstract}
  We extend our study of the extent of the regions within the $\alpha$ Centauri AB star system where small planets are able to orbit for billion-year timescales (Quarles \& Lissauer 2016, AJ 151, 111) to investigate the effects of minimizing the forced eccentricity of initial trajectories.  We find that initially prograde, circumstellar orbits require a piecewise quadratic function to accurately approximate forced eccentricity as a function of semimajor axis, but retrograde orbits can be modeled using a linear function.  Circumbinary orbits in the $\alpha$ Centauri AB system are less affected by the forced eccentricity.  {Planets on circumstellar orbits that begin with eccentricity vectors near their forced values are generally stable, up to $\sim$$10^9$ yr, out to a larger semimajor axis than are planets beginning on circular orbits. The amount by which the region of stability expands is much larger for retrograde orbits than it is for prograde orbits.  The location of the stability boundary for two planet systems on prograde, circular orbits is much more sensitive to the initial eccentricity state than it is for analogous single planet systems.} 
\end{abstract}





\section{Introduction}

The $\alpha$ Centauri star system contains the Solar System's nearest stellar neighbors.  The primary, $\alpha$ Centauri A, is slightly more massive than the Sun, the secondary,  $\alpha$ Centauri B, has slightly less than a solar mass, and the two stars orbit one another with a period of $\sim$80 years and an eccentricity just over 0.5.  If an Earth-like planet is present in the system, it could in principle be detected using a small space-based telescope \citep{Belikov2015b}.  The $\alpha$ Centauri system is billions of years old, so planets are only expected to be found in regions where their orbits are long-lived.  In \citet[henceforth referred to as Paper I]{Quarles2016}, we evaluated the extent of the regions within the $\alpha$ Centauri AB star system where small planets \textit{initially on circular orbits}, which in most cases were inclined relative to the plane of the binary orbit, are able to orbit for billion-year timescales;  Paper I also analyzed the stability of some planets with initial eccentricities, but only for circumstellar orbits in the plane of the binary and restricted to a single initial periapse longitude that was $\sim$75$^\circ$ from the periapse of the binary. \cite{Giuppone2017} have studied orbital stability in the $\alpha$ Centauri  system  through the chaos indicator MEGNO \citep{Cincotta2000} and found that regular orbits could exist at larger semimajor axes for eccentric orbits with other periapse longitudes.  

Because the binary companion induces a forced eccentricity upon the orbits of planets in orbit around either star, planets initially on circular orbits begin with nonzero free eccentricities \citep{Michtchenko2001,Michtchenko2004}.  The total eccentricities of such planets oscillate on timescales of order $\sim 10^4$ years, reaching peak values approximately twice as large as their forced eccentricity.  {Planets on circumstellar orbits in the eccentric binary star system $\gamma$ Cephei that have initial eccentricities equal to their forced eccentricity would experience far smaller oscillations than planets at the same semimajor axis beginning on circular orbits, and thus would not reach as high values of eccentricity \citep{Giuppone2011}.}  Therefore, we expect that planets on circumstellar orbits  in the $\alpha$ Centauri AB system with appropriately-phased initial eccentricities can be stable at a somewhat larger semimajor axis than are planets with initially circular orbits.

Our study uses updated values of the stellar masses and the configuration of the binary orbit given in the observational ephemeris derived by \cite{Pourbaix2016}, which we list in Table 1.  We include a brief comparison of the stability limits of particles on circular orbits using these new values {to compare} with the limits from our study in Paper I, which used stellar parameters from \cite{Pourbaix2002}.  Note that the current uncertainties in stellar parameters are comparable in magnitude to the differences between the 2002 values and those from 2016.

{Some studies of planetary stability in binary star systems have been performed starting the planets on circular orbits} (\citealt{Wiegert1997}; \citealt{Popova2012}; our Paper I).  {Other works \citep{Giuppone2011,AndradeInes2014,Rafikov2015b,Silsbee2015,Andrade2016,Andrade2017} have identified that choosing the planetary orbits relative to the architecture of the binary orbit may play a larger role in the formation and stability of planets within these systems. Additionally}, the first simulations that we performed in our study of the stability of multi-planet systems orbiting $\alpha$ Cen A or $\alpha$ Cen B showed that initially circular orbits lead to far less stable configurations than around single stars \citep{Quarles2017b}.  As a result, we were inspired to investigate these effects on the stability of small (Earth-mass) planets within $\alpha$ Centauri AB.

{We present herein the results of simulations that analyze planets starting near their forced eccentricities to quantify the effects on orbital stability.  Our methods and the initial conditions for our simulations are summarized in Section \ref{sec:methods}.  In Section \ref{sec:comp}, we compare the stability of particles on initially prograde orbits about $\alpha$ Centauri B using the model setup from Paper I with  updated system parameters and methods in this work.  We discuss forced eccentricities within the $\alpha$ Centauri AB system in Section \ref{sec:force_ecc}.  The results of our study for single-planet systems are presented in Section \ref{sec:singles}.  Results for an analogous study of the stability of two-planet systems are given in Section \ref{sec:multis}.  We provide the conclusions of our work and compare our results with previous studies in Section \ref{sec:conc}.}

\section{Methodology}\label{sec:methods}

The numerical simulations in this paper use the same custom version of the \texttt{mercury6} integration package designed to efficiently integrate orbits within a binary star system \citep{Chambers2002} that we employed in Paper I to evaluate the long-term stability of planetary orbits within the $\alpha$ Centauri system.  Similar to Paper I, the simulations assume standard Newtonian gravity and integrate each trajectory until a termination event occurs.  We define our termination events as a collision with either star, a body is ejected from the system, or a specified time interval elapses.  We consider planets initially on circumstellar orbits to be ejected if they reach radial distances larger than 50 AU from their host star and those on circumbinary orbits to be ejected if their distance from the center of mass of the system exceeds 200 AU.  

Our simulations model the planets as fully-interacting (massive) bodies with mass equal to that  of the Earth. Treating each planet individually (rather than as a large group of test particles) and using the hybrid symplectic method for integration with \texttt{mercury6} allows us to appropriately scale the choice of the starting timestep in terms of the Keplerian period (relative to the hosting star) while maintaining a tight control on the errors in energy and angular momentum.  

All of our simulations use the nominal values for the stellar masses and the configuration of the binary orbit given in the observational ephemeris derived by \cite{Pourbaix2016}, which we list in Table 1. We also report the uncertainties given in \cite{Pourbaix2016} for completeness.  Our simulations begin the stellar orbit at a mean anomaly of 209.6901$^\circ$, which corresponds to an epoch of JD 2452276.   

The planets reside within the binary plane in either prograde (inclination, $i_1 = 0^\circ$) or retrograde ($i_1=180^\circ$) motion relative to the binary orbit, where the planet's forced eccentricity is expected to be similar to or larger than those of analogous planets whose orbits are inclined to the binary plane \citep{Andrade2016}.  In our prograde runs, we begin the planet(s) with a longitude of periastron relative to the binary plane \citep{Michtchenko2001} defined by:
\begin{align}
\Delta \varpi \equiv \varpi_1 - \varpi_{bin},
\end{align}
the longitude of ascending node $\Omega$ is zero due to the near coplanarity, and mean anomaly $M$ determined for each planet using the relation with the golden ratio given in \cite{Smith2009}, where $M_1 = 222.492^\circ$ and $M_2 = 84.984^\circ$ for the first and second planet, respectively.  We also determine $\Delta \varpi$ numerically (See Section \ref{sec:ecc_contours}), as the previous works \citep{Andrade2016} make certain assumptions that would influence the value of $\Delta \varpi$ and find that its value varies with the orbital direction of the particle.  The conditions for the retrograde simulations are almost identical to the prograde simulations, except that their longitude of periastron is misaligned with the binary orbit such that $\Delta \varpi = -154^\circ$ {with respect to the plane of the sky, which corresponds to $\Delta \varpi = 180^\circ$ when measuring relative to the binary orbital plane. } 

We study planets in circumstellar orbits over a range of initial semimajor axes $a_1$ starting at 0.5 AU around either stellar component and extending to 4 AU (prograde) and 5.6 AU (retrograde).  Particles in circumbinary orbits cover a larger range of initial semimajor axes $a_1$ (40 -- 90 AU), as motivated from the results of Paper I.  

Dynamical effects, such a mean motion resonances, are expected to have their largest effects on system lifetime over small intervals in the planet semimajor axis near the stability limit.  Therefore, we take small steps (0.02) in the initial period ratio ($T_{bin}/T_j$) between the outermost planetary body and the companion star in order to resolve these small details.  Our other runs that investigate the full range of semimajor axes use a constant increment in semimajor axis ($\Delta a_1 = 0.005$ AU).  Resonant effects can occur over a broader range in semimajor axes in the circumbinary runs, where larger steps ($\Delta a_1$ = 0.1 AU) are taken in the planet's semimajor axis.

\section{Comparison with Paper I} \label{sec:comp}
\begin{figure*}
\centering
\includegraphics[width=\linewidth]{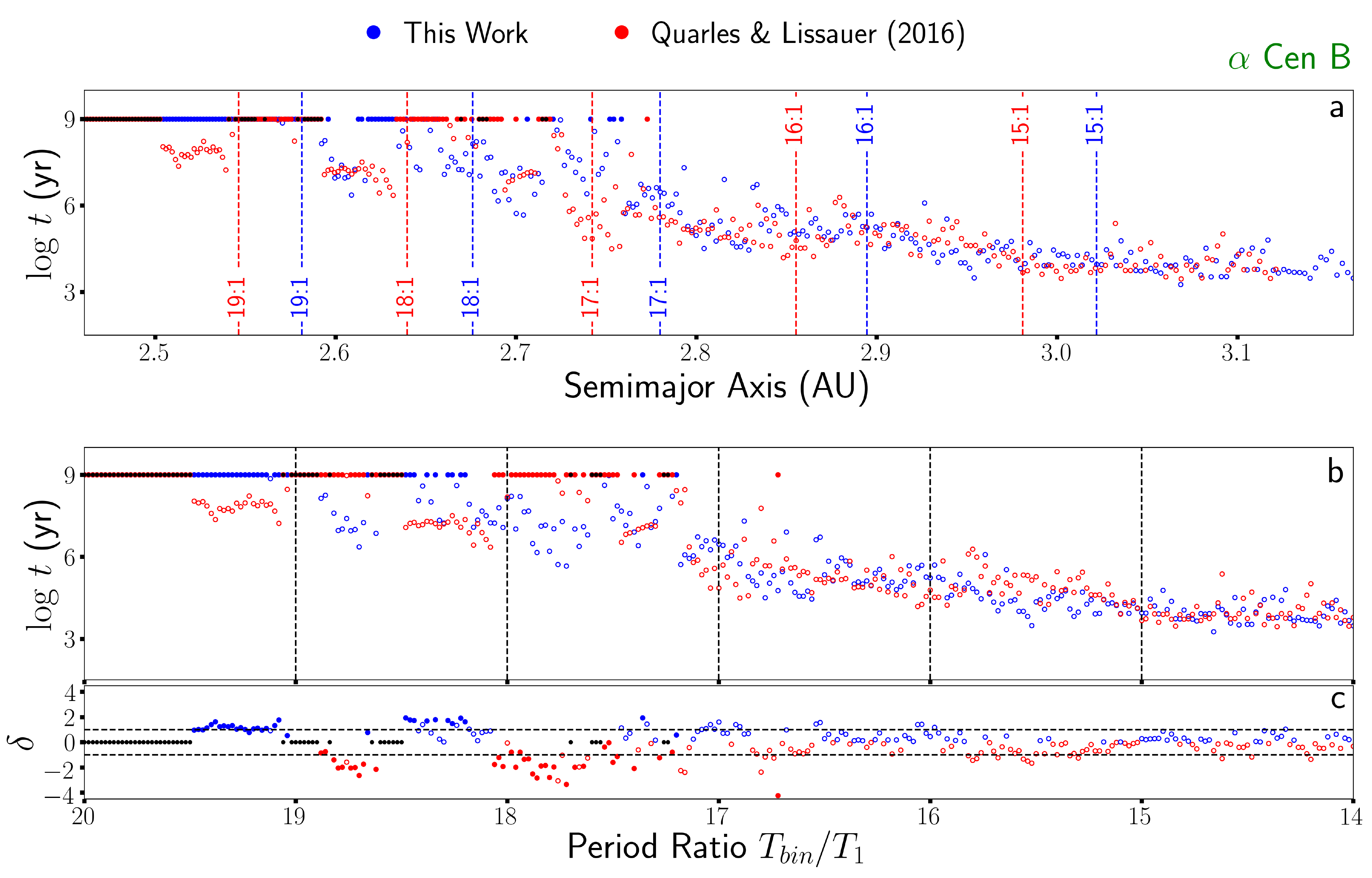}
\caption{Comparison between the lifetimes of particles on planar, prograde, initially circular orbits about $\alpha$ Cen B in the current work (blue) with those in Paper I (red). Filled circles indicate survival for the entire $10^9$ yr simulated, where the black circles denote survival for  $10^9$ yr  in both cases. Panel (a) illustrates the lifetimes as a function of the initial semimajor axis.  Panel (b) displays the results as a function of the initial period ratio between the binary and planetary orbits.  Panel (c) illustrates the difference in the logarithm of the lifetime with same initial period ratio between the two studies, $\delta$, where the points are color coded blue to signify that particles in the current work are longer lived, red meaning that particles in Paper I are longer lived, and black when both systems survive for the entire duration of the simulations.  The dashed vertical lines (panels a \& b) mark the locations of the $N:1$ mean motion resonances of the outer planet with the binary orbit and the dashed horizontal lines (panel c) mark when $\delta = \pm 1$, where points between these lines are within an order of magnitude.}
\label{fig:aCenB_comp}
\end{figure*}

In Paper I, we investigated the $\alpha$ Cen AB system using the best known parameters of the binary at the time \citep[e.g.,][]{Pourbaix2002} and used standard assumptions on the planetary orbit (i.e., $\omega = \Omega = M = 0^\circ$).  {We perform intermediate simulations with the initial planetary semimajor axis from 2.4 -- 3.2 AU and only update the binary orbit.  We find that this change shifts the instability regions near mean motion resonances outwards in planetary semimajor axis, but occur near the appropriate mean motion resonance.}  In this work, we have updated our methodology in choosing a different planetary longitude and a slightly different binary orbit (See Section \ref{sec:methods}).  We illustrate the effects of these changes by comparing the lifetimes of planets on prograde, planar, initially circular orbits around $\alpha$ Cen B.  {We determine the bulk stability of the ensemble of planets in this region as 27.2\%, 21.6\%, and 26.9\% for the simulations from Paper I, our intermediate runs, and our new simulations presented in Figure \ref{fig:aCenB_comp}, respectively.  Note that while the differences in these values are useful for comparing the effective limits of the stability zones for these three cases, they would all go up or down dependent on the assumed window of planetary semimajor axis values considered.}

Figure \ref{fig:aCenB_comp} shows the contrast between { the methodology from Paper I and our current one that updates the binary orbit and uses a different initial planetary longitude.} The locations of the mean motion resonances are displaced due to the updated masses of the stars, so we compare the lifetimes of planets at a similar semimajor axis in Fig. \ref{fig:aCenB_comp}a, and those at similar period ratio in Fig. \ref{fig:aCenB_comp}b.   The stability limit is slightly farther out ($\sim$2.59 AU) compared to the previously determined value (2.54 AU) as shown in Fig. \ref{fig:aCenB_comp}a{, where in both cases the stability limit is near the 19:1 mean motion resonance.}  This is partly due to the shift of the $N:1$ mean motion resonances (Fig. \ref{fig:aCenB_comp}b), where particles were typically stable at period ratios larger than 19.  Also some `lucky' orbits may not appear anymore due to a different initial phase of the planet. Figure \ref{fig:aCenB_comp}c shows that only a small fraction of initial conditions lead to substantially different outcomes (i.e., larger than an order of magnitude), indicating that the two works are in rough agreement in the region close to the stability limit.
 
\section{Forced Eccentricities} \label{sec:force_ecc}
\subsection{Identifying the Forced Eccentricity}\label{sec:ecc_contours}
\begin{figure*}
\epsscale{1.}
\centering
\plotone{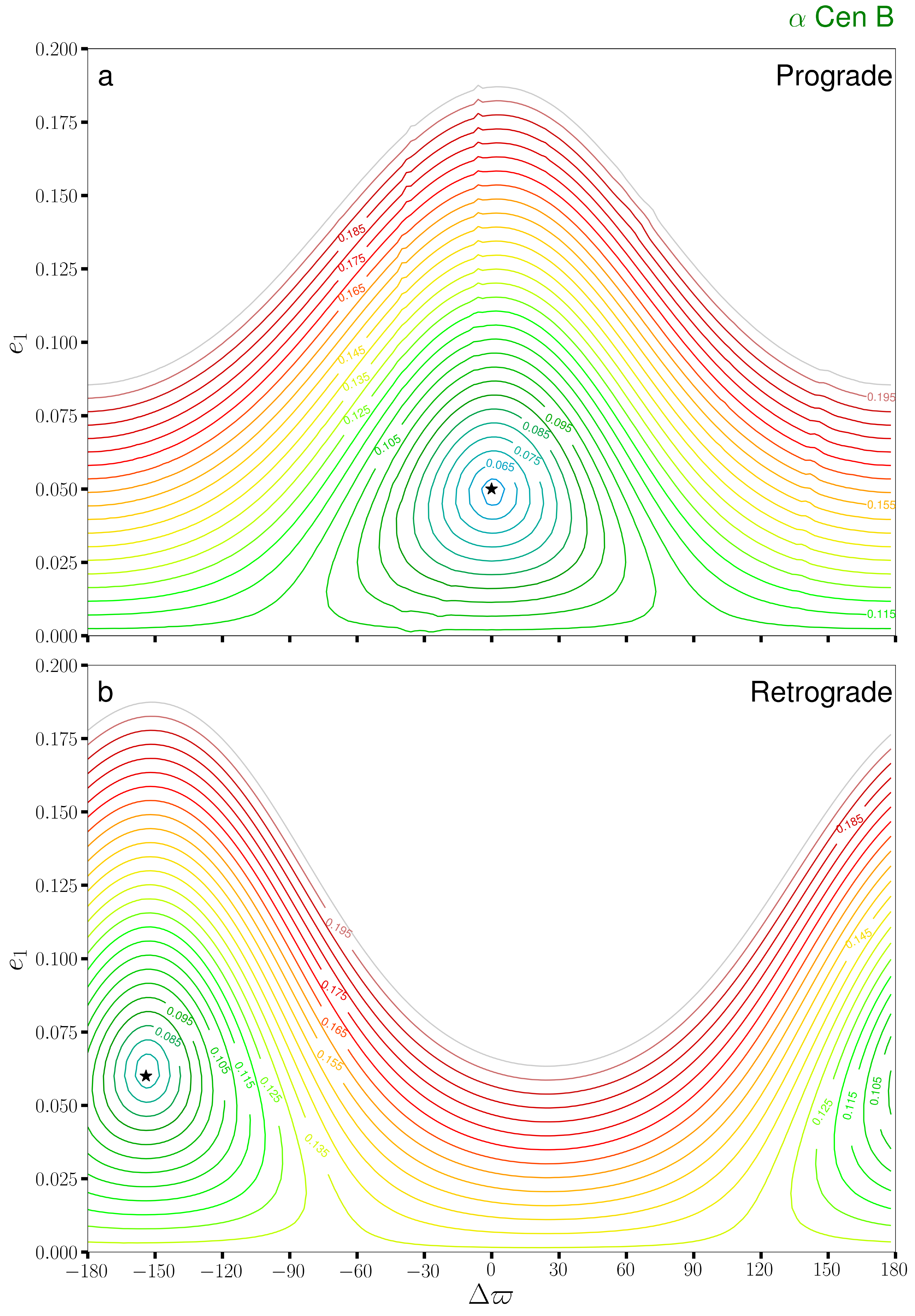}
\caption{Contours of maximum eccentricity derived from numerical simulations over 100,000 years for a planet initially orbiting $\alpha$ Cen B with a semimajor axis of 1.5 AU. The simulations are performed over a grid in the initial eccentricity $e_1$ and the relative argument of periastron $\Delta \varpi$ to determine the appropriate values of $\Delta \varpi$ that minimize free eccentricity for a prograde (a) or retrograde (b) orbit.}
\label{fig:ew_aCenB}
\end{figure*}

We present example simulations in Figure \ref{fig:ew_aCenB} to demonstrate how the maximum eccentricity varies for an Earth-mass planet of semimajor axis $a_1 = 1.5$ AU within the phase space of its initial eccentricity, $e_1$, and longitude of periastron relative to the binary orbit, $\Delta \varpi$ (see Section \ref{sec:methods}).  Each panel in Fig. \ref{fig:ew_aCenB} uses results from a uniform grid of simulations in $e_1 - \Delta \varpi$ space resulting in a resolution of $101\times180$ initial conditions ($0 \le e_1 \le 0.2$ in steps of 0.002 and $2^\circ$ increments in $\Delta \varpi$) over a timescale of 100,000 years.  The contours are drawn using the maximum eccentricity achieved over a given simulation, where we include contour levels only up to $e_{max} = 0.2$ in steps of 0.005.  In this way, we can define numerically the forced eccentricity vector as the initial conditions that produces the lowest maximum eccentricity measured, which is similar to the definition used by \cite{Andrade2016} (their Section 3.1).  

Figure \ref{fig:ew_aCenB}a shows that a minimum occurs for $e_1 = 0.05$ and $\Delta \varpi = 0^\circ$ for a prograde orbit, and Fig. \ref{fig:ew_aCenB}b reveals that the minimum for a retrograde orbit occurs when $e_1 = 0.065$ and $\Delta \varpi = -154^\circ$.  If we perform similar integrations at larger and smaller semimajor axes of the planet, the minimum value of $e$ moves up or down in response to a greater or lower perturbation, respectively.  However, the value of $\Delta \varpi$ remains unchanged as it is set by the binary orbit, which we keep fixed among all of our simulations.

\begin{figure*}
\epsscale{1.}
\centering
\plotone{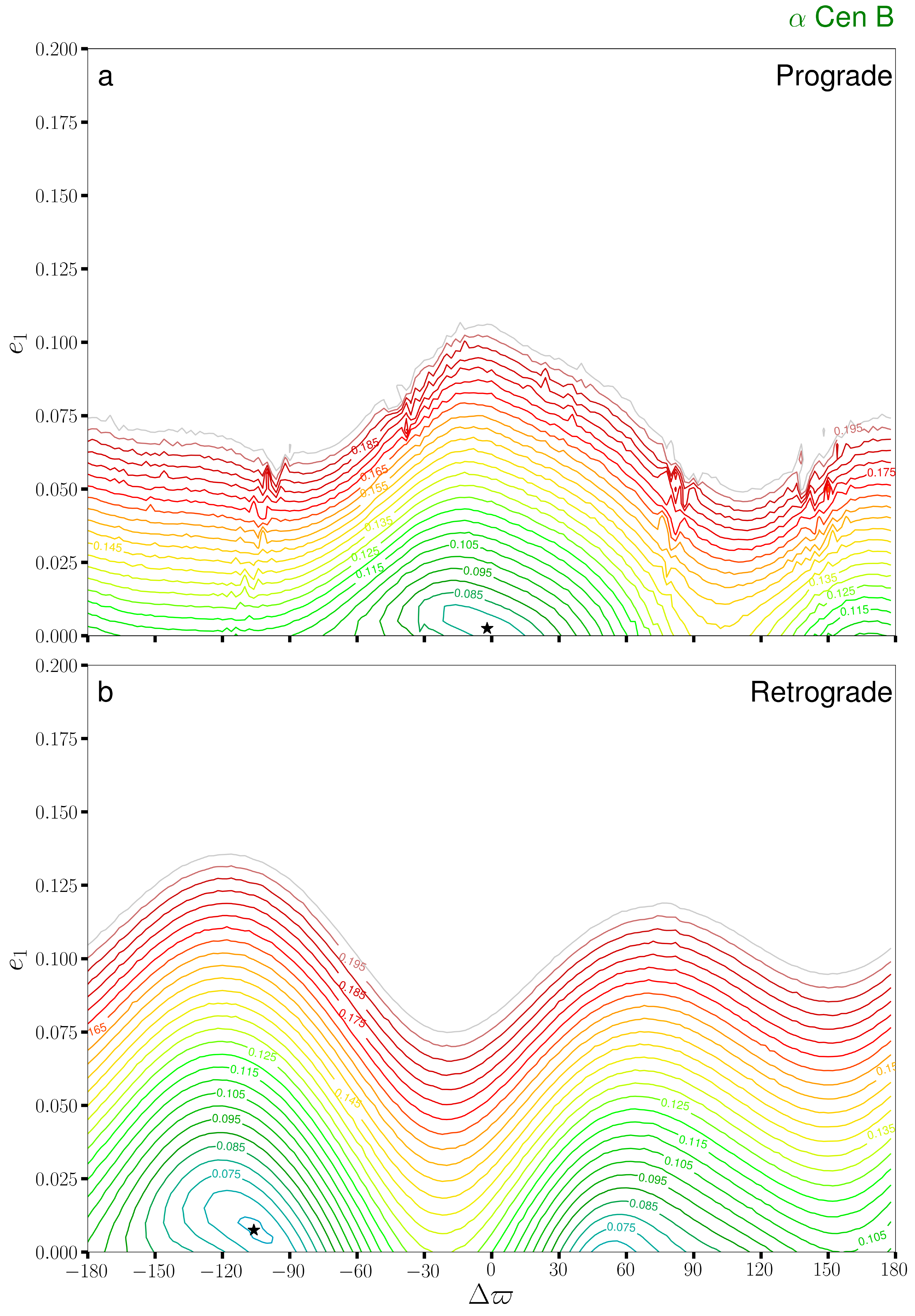}
\caption{Contours of maximum eccentricity derived from numerical simulations over 100,000 years for a planet initially orbiting both stars with a semimajor axis of 85 AU. The simulations were performed over a grid in the initial eccentricity, $e_1$, and the relative argument of periastron, $\Delta \varpi$. The stars represent locations of the minimal values of the peak total eccentricity reached, and therefore provide good estimates of the forced eccentricity for a prograde (a) or retrograde (b) orbit. The upper curves in panel (a) are choppy because planets at 85 AU with the represented eccentricities are near the stability boundary for $10^5$ year integrations.}
\label{fig:ew_aCenAB}
\end{figure*}

We perform a similar analysis for circumbinary orbits, where we chose $a_1 = 85$ AU informed by Paper I.  Figure \ref{fig:ew_aCenAB} illustrates these results, with prograde orbits in Fig. \ref{fig:ew_aCenAB}a and retrograde orbits in Fig. \ref{fig:ew_aCenAB}b.  There is a stark contrast between Figs. \ref{fig:ew_aCenB} and \ref{fig:ew_aCenAB}, where the latter indicates initially circular orbits to have the lowest value in maximum eccentricity.  

\subsection{Eccentricity Variations} \label{sec:eccvec}
\begin{figure*}
\epsscale{1.}
\centering
\plotone{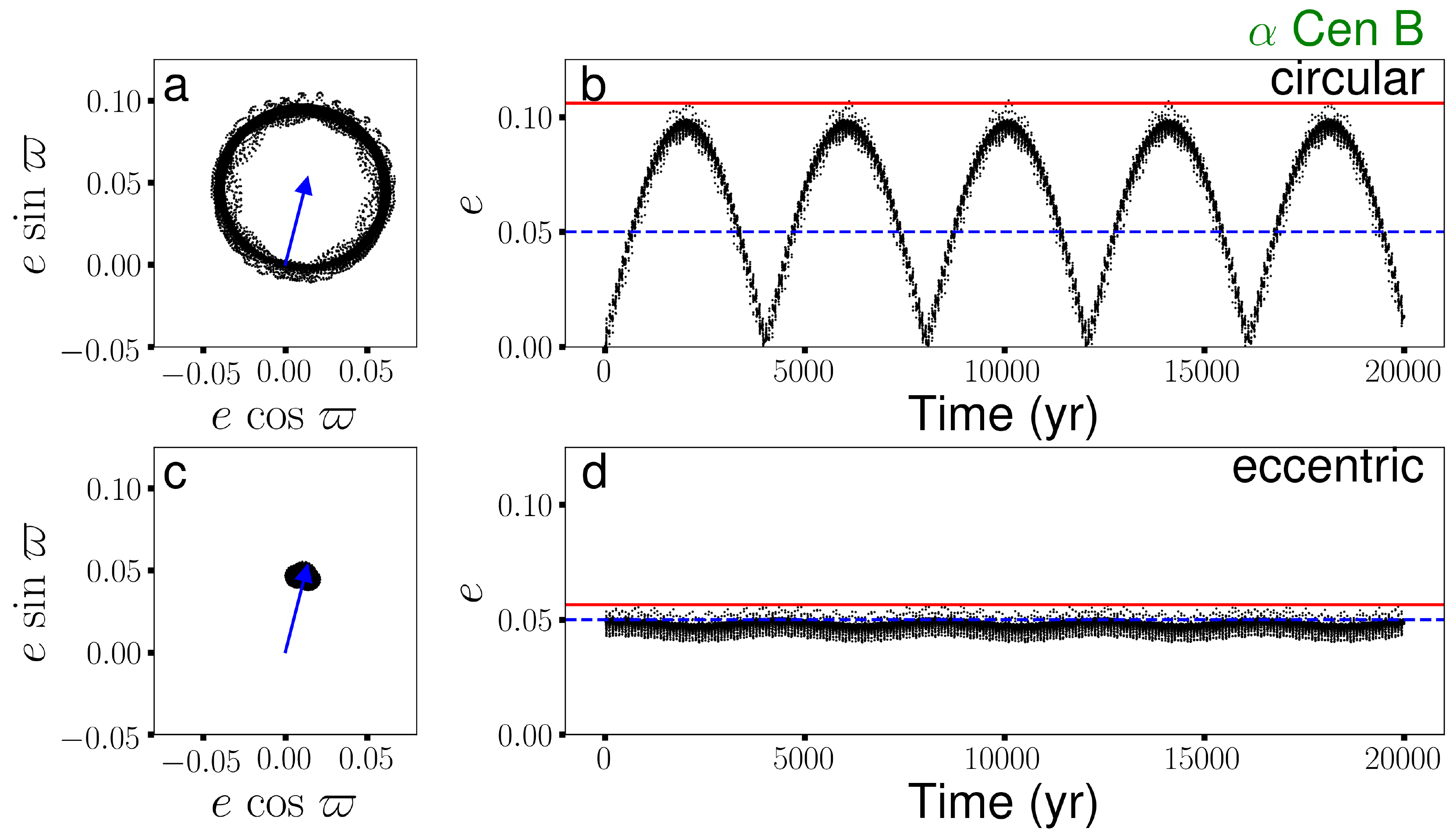}
\caption{Components of the eccentricity vectors (a \& c) and the magnitude of eccentricity for 20,000 years of evolution for a {prograde} planet initially orbiting $\alpha$ Cen B with a semimajor axis of 1.5 AU.  The planet represented on the top row (a \& b) begins on a circular orbit, and the planet represented on the bottom row (c \& d) begins near the forced eccentricity due to $\alpha$ Cen A.  The blue arrows and blue dashed lines correspond to the forced components of eccentricity.  The red horizontal lines mark the maximum eccentricity reached in the first 5,000 years of evolution.}
\label{fig:ecc_aCenB_Pro}
\end{figure*}

Figure \ref{fig:ecc_aCenB_Pro} displays the evolution of the total eccentricity vector of  planets on prograde orbits with $a_1 = 1.5$ AU and initial eccentricity either equal to zero or near the forced eccentricity over 20,000 years (a few secular cycles). Panels \ref{fig:ecc_aCenB_Pro}b and \ref{fig:ecc_aCenB_Pro}d show the time series evolution of the osculating eccentricity of the planet.  The blue dashed horizontal line in these figures marks the magnitude of the forced eccentricity, and the red solid horizontal line marks the maximum value of eccentricity encountered within the first 5,000 years.  The difference between the maximum eccentricity in Figs. \ref{fig:ecc_aCenB_Pro}b and \ref{fig:ecc_aCenB_Pro}d is non-trivial.  Moreover, starting the planet at a semimajor axis near the stability limit increases this difference in eccentricity variation.

\begin{figure*}
\epsscale{1.}
\centering
\plotone{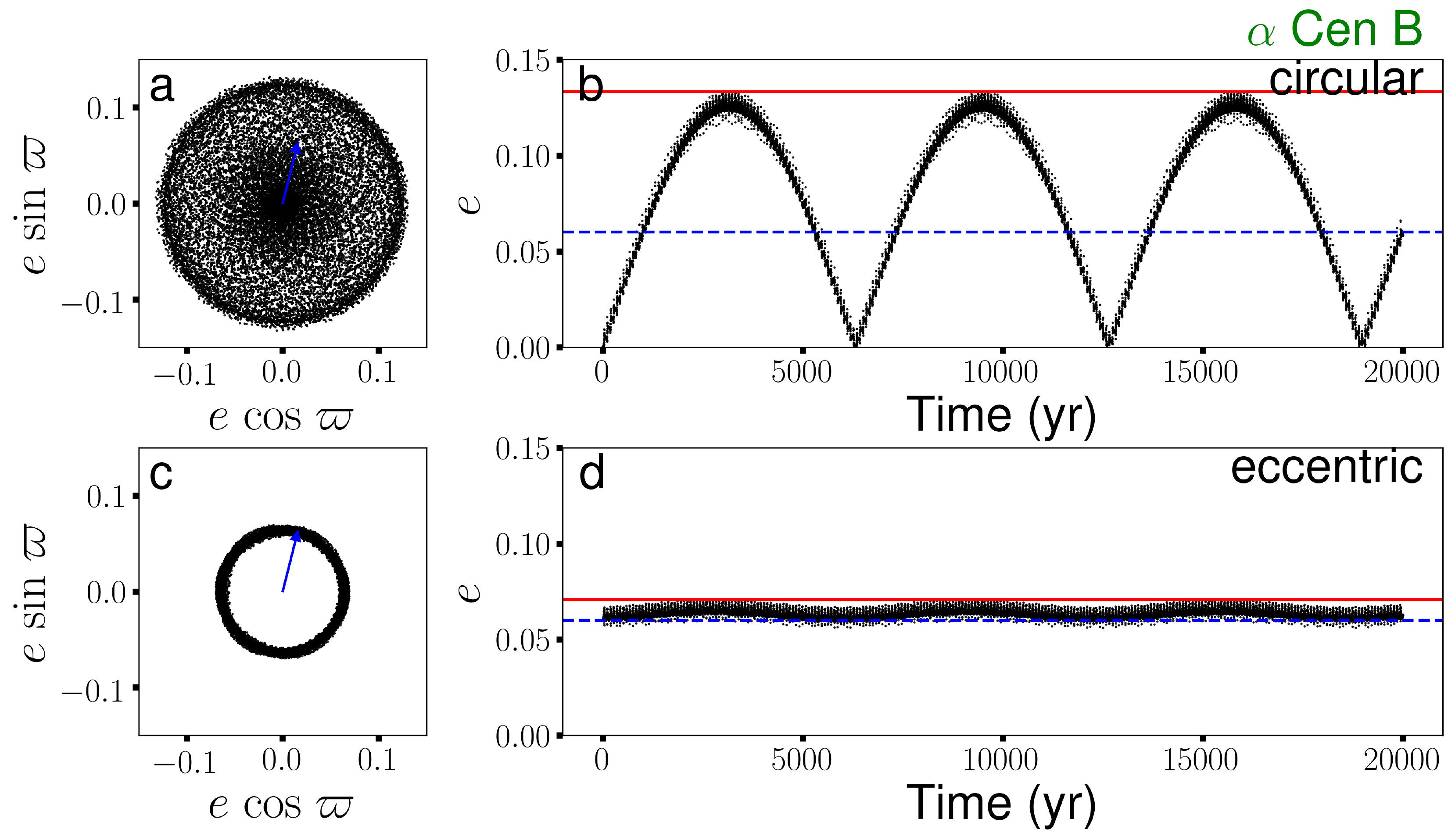}
\caption{Components of the eccentricity vectors (a \& c) and the magnitude of eccentricity for 20,000 years of evolution for a {retrograde} planet initially orbiting $\alpha$ Cen B with a semimajor axis of 1.5 AU.  The planet represented on the top row (a \& b) begins on a circular orbit, and the planet represented on the bottom row (c \& d) begins near the forced eccentricity due to $\alpha$ Cen A.  The blue arrows and blue dashed lines correspond to the forced components of eccentricity.  The red horizontal lines mark the maximum eccentricity reached in the first 5,000 years of evolution.}
\label{fig:ecc_aCenB_Ret}
\end{figure*}

We perform a similar set of simulations for retrograde orbits using the appropriate components of the forced eccentricity vector derived from Fig. \ref{fig:ew_aCenB}b.  These retrograde simulations, whose results are shown  in Figure \ref{fig:ecc_aCenB_Ret}, exhibit a similar trend in lowering the maximum eccentricity when starting the planet with an appropriately phased eccentricity.   

\subsection{Formulas for Forced Eccentricity}
Recently, \cite{Andrade2016} {and \cite{Andrade2017}} investigated how the forced eccentricity varies within the context of a restricted three-body problem using first- and second-order perturbation theories.  We are exploring a similar problem with planets in $\alpha$ Centauri AB, specifically with 2 circumstellar configurations.  Their numerical procedure for determining the forced eccentricity makes use of techniques based upon frequency analysis on numerical integration \citep{Laskar1990,Michtchenko2002}, where a time averaged component of the eccentricity vector is determined iteratively.  We follow a similar iterative process using our values for $\Delta \varpi$ determined in Section \ref{sec:ecc_contours} in a series of numerical simulations \citep{Noyelles2008,Couetdic2010}.

\begin{figure*}
\epsscale{1.}
\centering
\plotone{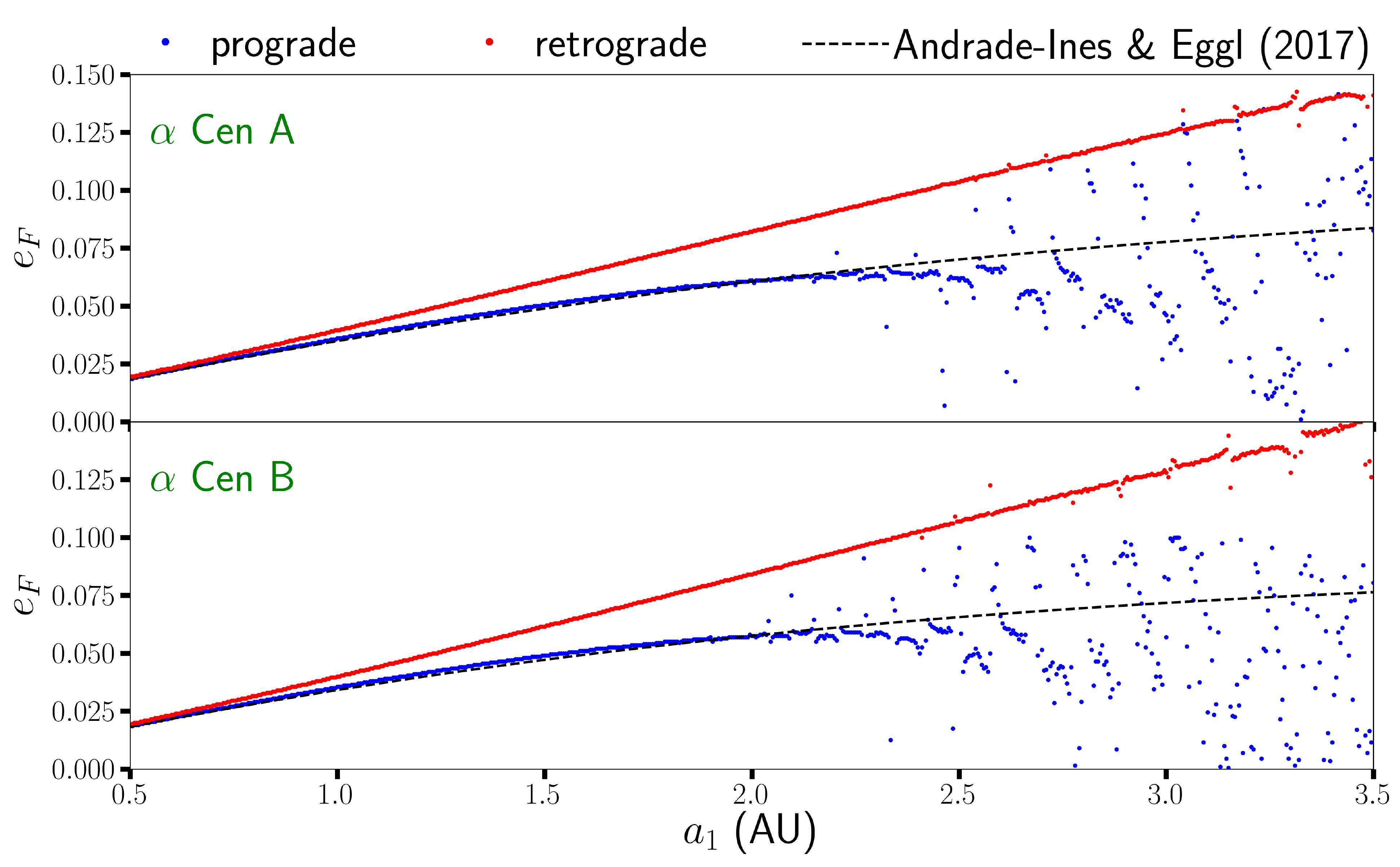}
\caption{Map of the forced eccentricity, $e_F$, relative to the initial semimajor axis, $a_1$, for a planet initially in orbit around $\alpha$ Cen A (a) or $\alpha$ Cen B (b) in either prograde (blue) or retrograde (red) orbits.  {The overplotted dashed black curve represents an appropriate approximation using higher-order secular theory \citep{Andrade2017}.} }
\label{fig:ecc_vec}
\end{figure*}

Our simulations seek to identify the maximum eccentricity through a grid of initial conditions over a range of the planet's initial eccentricity $e_1$ (0.0--0.20) in steps of 0.0005 and initial semimajor axis $a_1$ (0.5--3.5) in steps of 0.005 AU and using the relative {longitude of periastron} values determined from Sec. \ref{sec:ecc_contours}.  Following from Figure \ref{fig:ew_aCenB}, we identify the forced eccentricity $e_F$ as the lowest magnitude of the maximum eccentricity at a given planet semimajor axis $a_1$.  Figure \ref{fig:ecc_vec} shows how the forced eccentricity varies with $a_1$ for circumstellar planets orbiting each star in both the prograde (blue) or retrograde (red) directions.  \cite{Andrade2016} adopted an expansion to second order in perturbation theory, where we have implemented a hybrid approach between numerical and analytic methods.  {\cite{Andrade2017} developed a higher-order scheme to estimate the forced eccentricity of planets on prograde orbits, which is plotted as a dashed curve in Fig. \ref{fig:ecc_vec}. While it matches our numerical results nicely at small semimajor axes, it tends to overestimate the forced eccentricity relative to our numerical simulations for $a_1 > 2$ AU.} 

We fit the blue curves in Fig. \ref{fig:ecc_vec} using a quadratic function of the form:
\begin{align}
e_F &= C_1a_1^2 + C_2a_1 + C_3,
\end{align}
where $C_1$, $C_2$, and $C_3$ are our fitted coefficients.  A single function can be fit for semimajor axes less than 2 AU, but tend to overestimate the forced eccentricity at larger semimajor axes.  Thus, we split the parameter space and fit coefficients piece-wise using 2 AU as a break point.  The results of our fitting are given in Table \ref{tab:coeff_pro}.  The forced eccentricity behaves much more linearly for the retrograde cases, so we implement the following equation that has a form similar to \cite{Andrade2016} which uses the disturbing function from \cite{Heppenheimer1978}:

\begin{align}
e_F &= C_1\frac{a_1}{a_{bin}}\frac{e_{bin}}{1 - e_{bin}^2},
\end{align}

\noindent where $C_1$ is a fitted coefficient from our simulations, $a_{bin}$ denotes the binary semimajor axis, and $e_{bin}$ marks the eccentricity of the binary orbit.  The results of our fitting are given in Table \ref{tab:coeff_ret}.

\section{Regions Where a Single Planet Can Remain Stable}\label{sec:singles}
Having deduced an approximate value for the forced eccentricity vector (see Section \ref{sec:eccvec}), we investigate the differences in stability between planets initially on circular orbits and those starting with the forced eccentricity vector.  We focus on the region near the stability boundary that we found in Paper I for nearly coplanar, circular orbits . As shown in Paper I, $N:1$ MMRs destabilize specific regions, and thus we proceed with steps of 0.02 in the initial period ratio between the secondary star and the planet.  By choosing our steps in the $x$-axis in this way, we can more uniformly sample regions where the $N:1$ MMRs would be active.

\begin{figure*}
\epsscale{1.}
\centering
\plotone{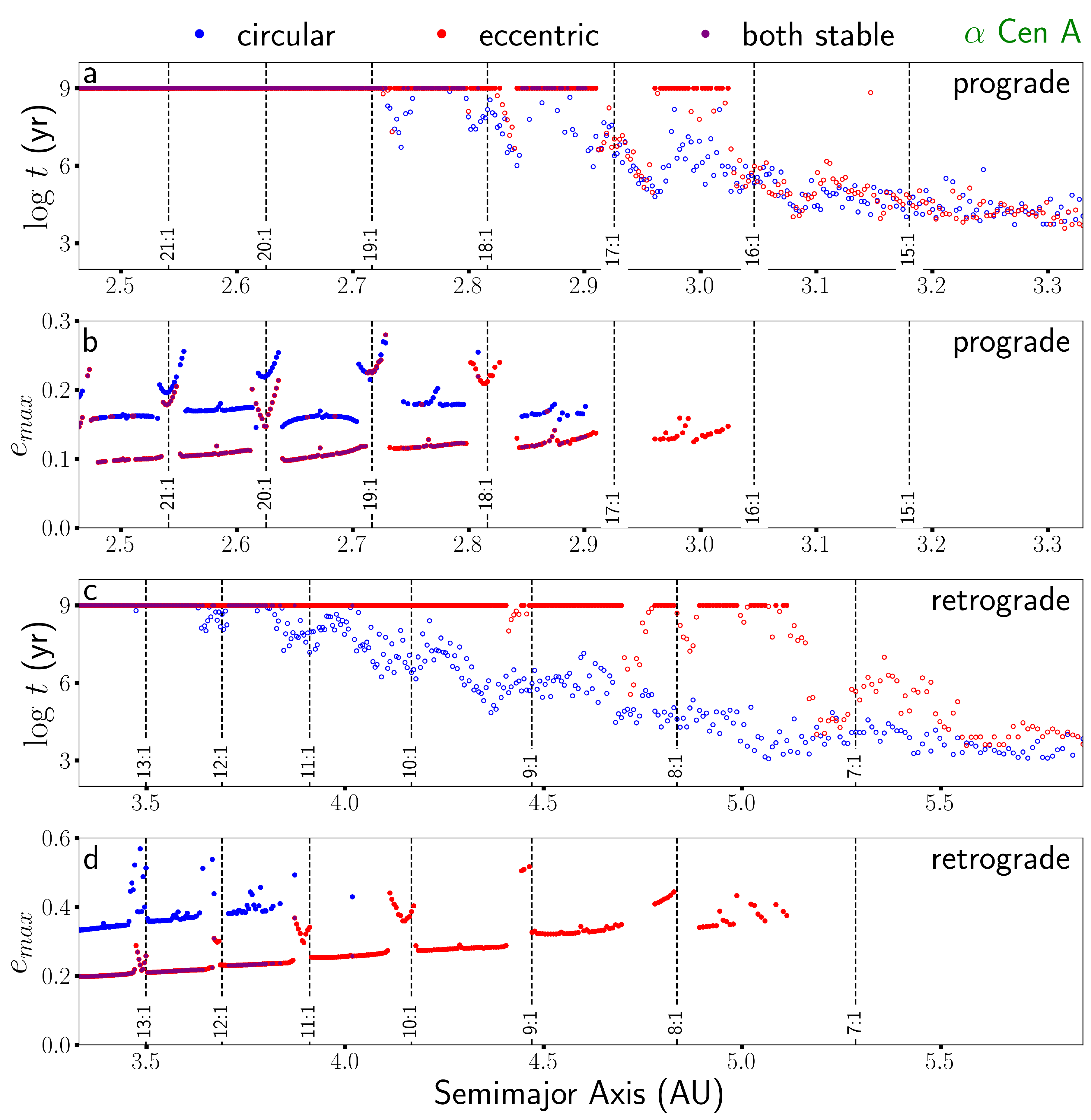}
\caption{Lifetime (a \& c) and maximum eccentricity (b \& d) of prograde and retrograde planets orbiting $\alpha$ Cen A as a function of the starting planetary semimajor axis.  Stable ($> 1$ Gyr lifetime) runs with initially circular (blue) or eccentric (red) orbits are filled, where unstable runs are open circles carrying the respective color-code.  Runs that are stable for both circular and eccentric orbits are overplotted in purple.  The vertical dashed lines and labels mark the locations of $N:1$ mean motion resonances with the binary orbit.}
\label{fig:life_aCenA}
\end{figure*}

Figure \ref{fig:life_aCenA} shows the results of our simulations for both initially prograde and retrograde planets around $\alpha$ Cen A, respectively.  In Fig. \ref{fig:life_aCenA}a, we mark the lifetime of a prograde planet as a function of the starting semimajor axis (converted from period ratio). All planets with $a_1 < 2.7$ AU are stable. We find substantial overlap between the survival times of planets on initially circular orbits (blue) with those beginning with the forced eccentricity (red) for semimajor axis values  $>$ 3.05 AU, where these are both unstable as opposed to both stable.  The region 2.7 -- 3.05 AU (equivalently, between the 16:1 and 19:1 MMRs) is where starting with the forced eccentricity vector matters the most (i.e., allowing for more  initial conditions at larger semimajor axes to be stable).  

We plot the maximum eccentricity of the stable configurations attained over the 1 Gyr simulated in Figure \ref{fig:life_aCenA}b.  From this view, we see that the initially circular orbits typically rise to higher values of eccentricity.  Moreover, the elevated eccentricities due to interactions with both the $N:1$ and $N:2$ MMRs with the binary are readily apparent.

For retrograde orbits, we examine much larger distances from the host star, $\sim$3.4 -- 6.0 AU.  Figure \ref{fig:life_aCenA}c illustrates that the lifetimes of the planets are strikingly different between our two cases, circular or eccentric, where  retrograde planets starting near their forced eccentricities can remain stable on a 1 Gyr timescale at a semimajor axis of $\sim$5 AU.   Elevated levels of the maximum eccentricity appear displaced from our marked locations for the $N:1$ MMRs (Fig. \ref{fig:life_aCenA}d).

\begin{figure*}
\epsscale{1.}
\centering
\plotone{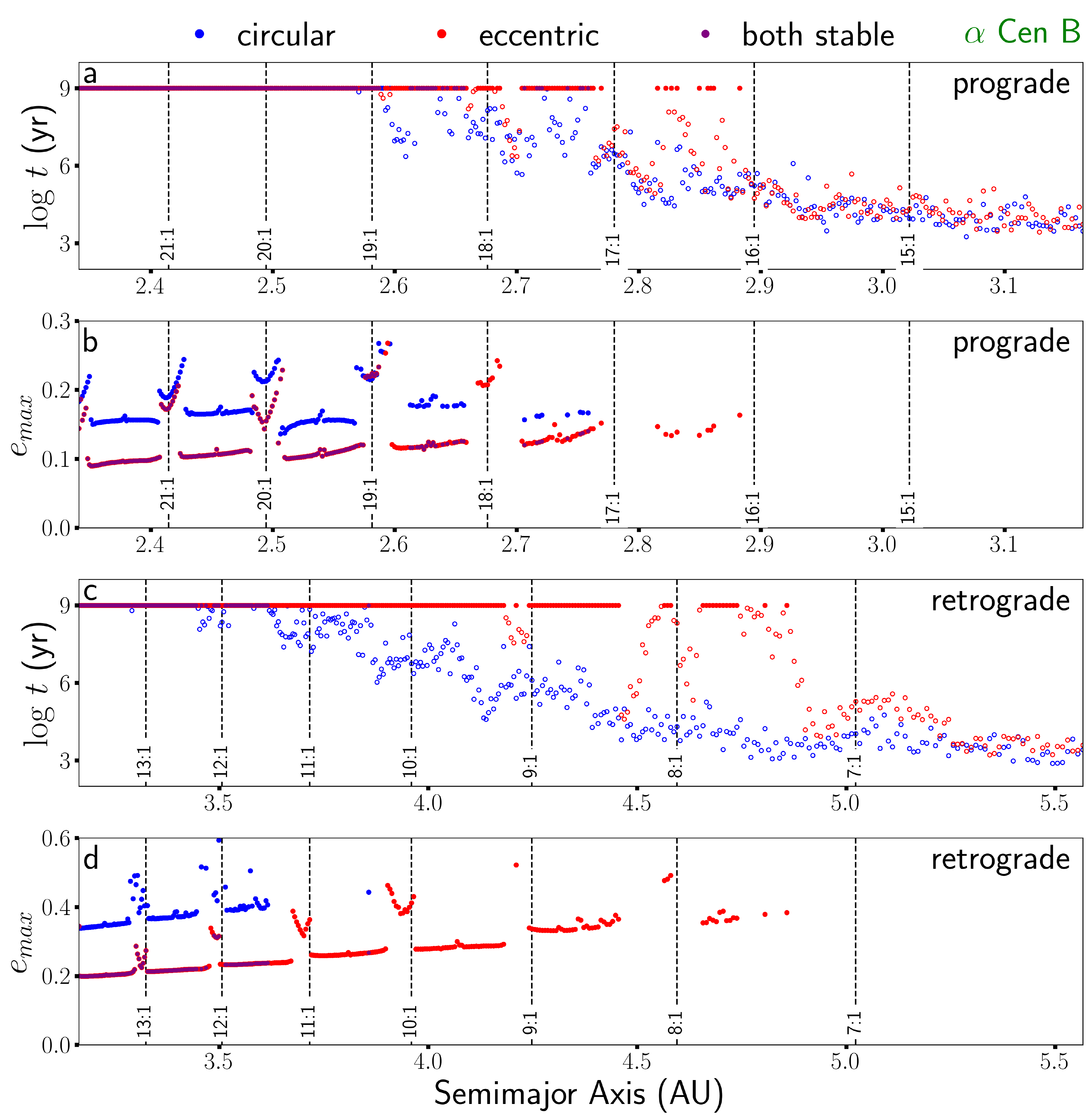}
\caption{Lifetime (a \& c) and maximum eccentricity (b \& d) of prograde and retrograde planets orbiting $\alpha$ Cen B as a function of the starting planetary semimajor axis.  Stable ($> 1$ Gyr lifetime) runs with initially circular (blue) or eccentric (red) orbits are filled, where unstable runs are open circles carrying the respective color-code.  Runs that are stable for both circular and eccentric orbits are overplotted in purple.  The vertical dashed lines and labels mark the locations of $N:1$ mean motion resonances with the binary orbit.}
\label{fig:life_aCenB}
\end{figure*}

For completeness, we show results when the planet begins in a prograde and retrograde orbit around $\alpha$ Cen B in Figure \ref{fig:life_aCenB}.  Due to the symmetry of the problem, we find similar outcomes, where the borders of stability (Figs. \ref{fig:life_aCenB}a and \ref{fig:life_aCenB}c) appear near similar MMRs and shift to smaller semimajor axis values.  

\begin{figure*}
\epsscale{1.}
\centering
\plotone{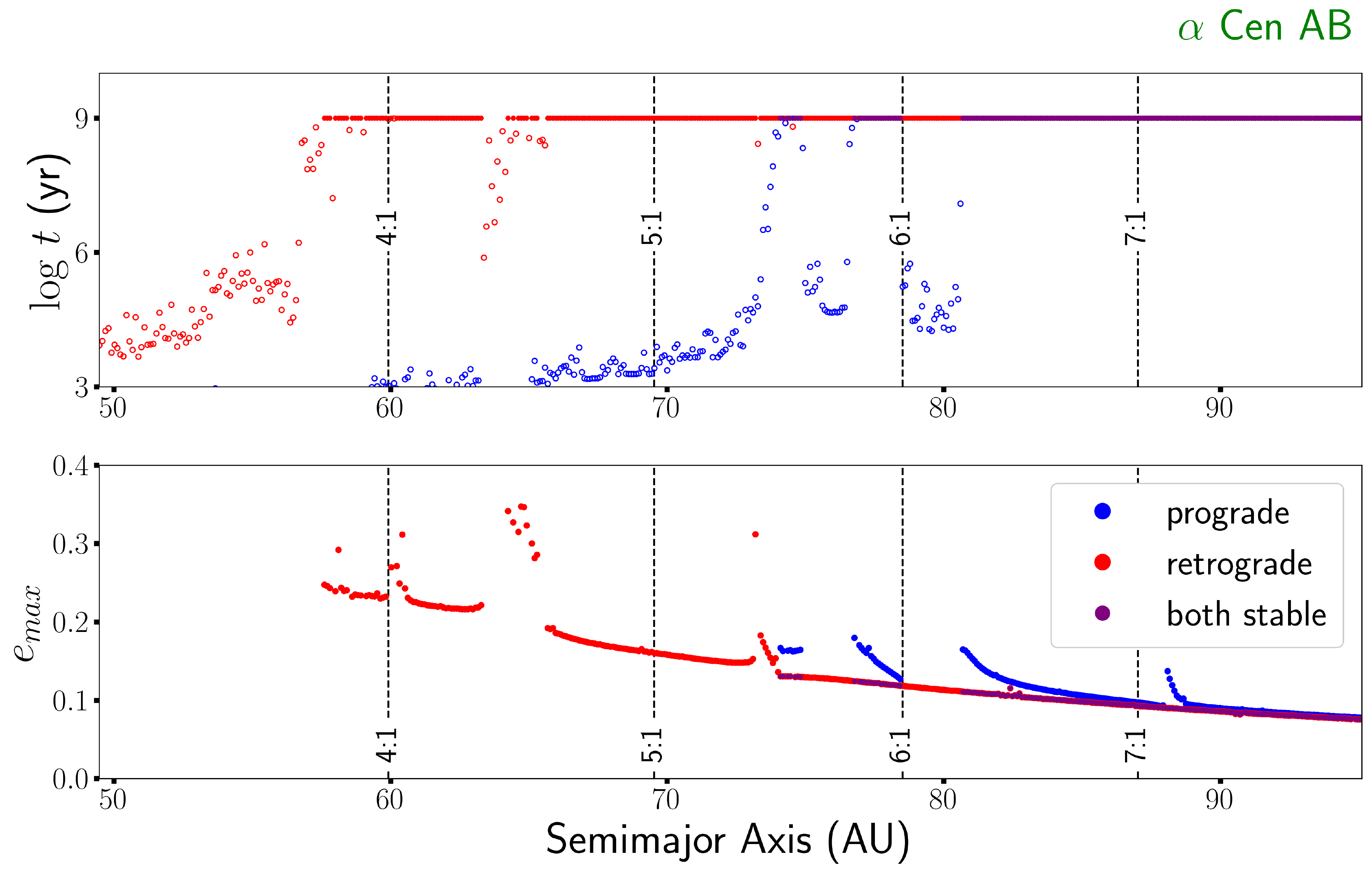}
\caption{Lifetime (top) and maximum eccentricity (bottom) of planets in circumbinary orbits around $\alpha$ Cen AB as a function of the starting planetary semimajor axis.  Stable ($> 1$ Gyr lifetime) runs with initially prograde (blue) or retrograde (red) orbits are filled, where unstable runs are open circles carrying the respective color-code.  Runs that are stable for both prograde and retrograde orbits are overplotted in purple.  The vertical dashed lines and labels mark the locations of $N:1$ mean motion resonances with the binary orbit.}
\label{fig:life_aCenAB}
\end{figure*}

Section \ref{sec:eccvec} shows that circumbinary orbits are not affected strongly by the assumption on the initial eccentricity, but are dependent on the initial phase relative to the binary orbit (see Fig. \ref{fig:ew_aCenAB}).  Thus, we provide results of circumbinary planets in which the respective initial longitude is modified between the prograde or retrograde simulations.  Figure \ref{fig:life_aCenAB} (top panel) illustrates that retrograde planets can stably orbit $\sim$20 AU closer to the binary than in the case of prograde.  These orbits are typically more eccentric than in the circumstellar case and likely allowed due to the initial phase difference with the binary orbit.  The locations of stable particles are slightly displaced compared to our results in Paper I, which is mainly a result of the orbital solution of the binary used in this work.

\section{Stability of Two-Planet Systems around $\alpha$ Cen A}\label{sec:multis}
We briefly investigate the lifetime and maximum eccentricity of the outermost planet in prograde two-planet systems orbiting $\alpha$ Cen A.  Our simulations follow a similar approach as in Section \ref{sec:singles} in terms of the steps in the semimajor axis $a_2$.  The other planet is placed interior to $a_2$ using the formalism developed in \cite{Smith2009}.  The difference between the initial semimajor axes of the two planets is taken to be {$12 R_{H,m}$, where the mutual Hill Radius of the two planets is given by \citep{Chambers1996},} 
\begin{equation}
R_{H,m} = {(a_1 + a_{2}) \over 2}\left[{m_1 + m_{2} \over 3(M_\star+m_1)}\right]^{1/3}.
\end{equation}
We explore a wider range of planet spacings in \cite{Quarles2017b}.  The planets either both have initially circular orbits or both begin  with the appropriate forced eccentricity (eq. 1).

\begin{figure*}
\epsscale{1.}
\centering
\plotone{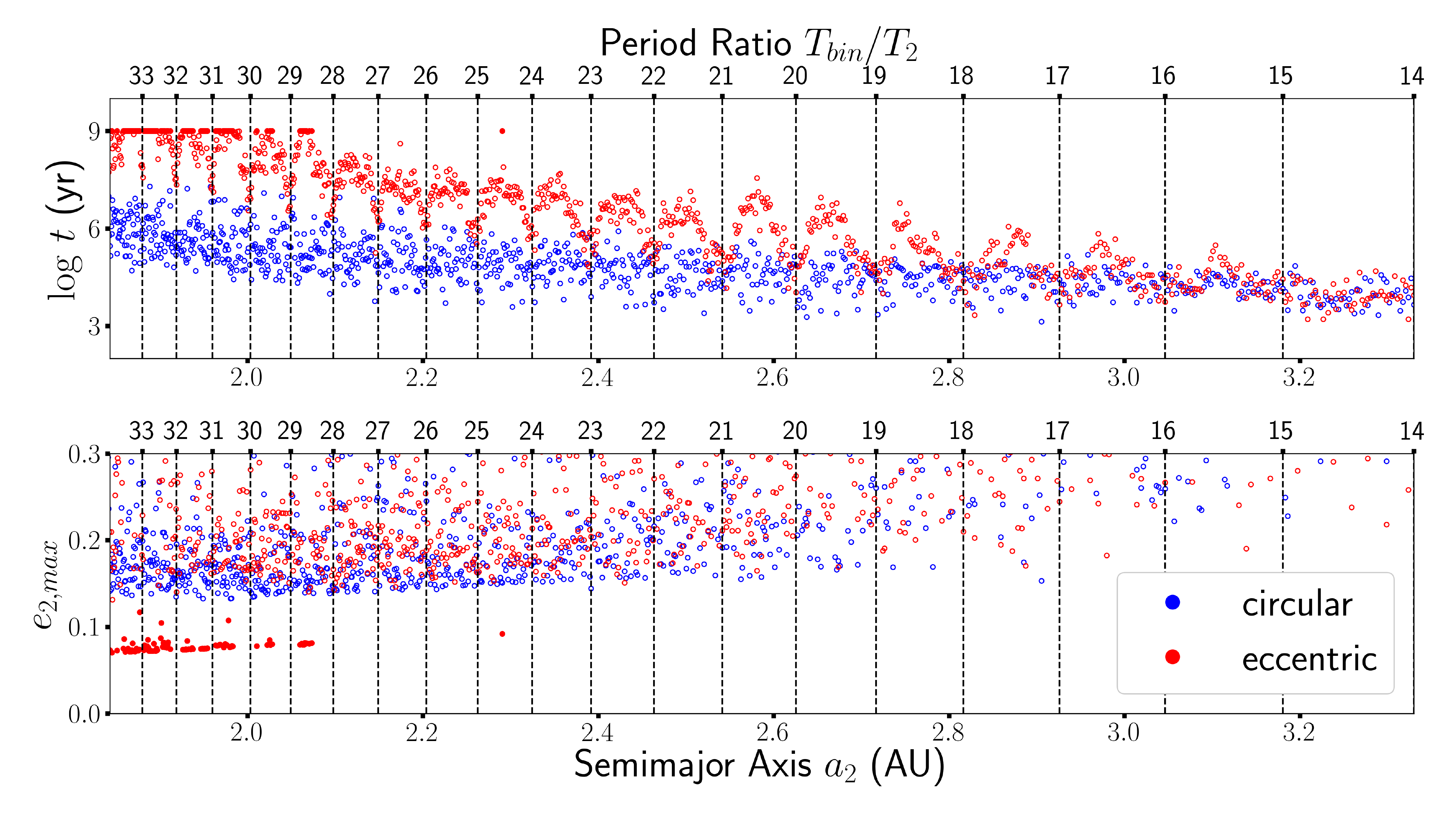}
\caption{Lifetime (top) and maximum eccentricity (bottom) of prograde two-planet systems orbiting $\alpha$ Cen A as a function of the starting planetary semimajor axis of the outermost planet, where the planets are separated by $12 R_{H,m}$ (see Eq. 4).  Stable ($> 1$ Gyr lifetime) runs with initially circular (blue) or eccentric (red) orbits are filled, where unstable runs are open circles carrying the respective color-code.  We note that the termination event is usually a collision between the two planets and the maximum eccentricity does not have to be very large for our definition of instability to occur.  The top $x$-axis labels and vertical dashed lines mark the locations of $N:1$ mean motion resonances of the outer planet with the binary orbit.  Note that all of the stable systems have $e_{2,max} < 0.15$.}
\label{fig:life_aCenA_2pl}
\end{figure*}

Figure \ref{fig:life_aCenA_2pl} (top panel) shows the systems' lifetimes, i.e.,  when a termination event (ejection or collision) occurs for either of the two planets.    The growth of eccentricity when the planets begin on circular orbits typically leads to a collision between the planets on a short ($\lesssim 10$ Myr) timescale, although it is possible for collisions with the stars or ejection (radial distance $> 100$ AU) from the system to occur.  We continued our circular runs inward (decreasing $a_2$) and found that stable configurations were possible for $a_2 \sim$1 AU \citep{Quarles2017b}. The lifetimes of two planets starting near their forced eccentricities is strongly correlated with proximity to the binary $N:1$ MMRs, which act as a destabilizing force.  Broad regions of stability are seen for values of $a_2 \lesssim 2.1$ AU.

Figure \ref{fig:life_aCenA_2pl} (bottom panel) shows the maximum eccentricity of the outermost planet, $e_{2,max}$, up to the termination event.  This contrasts sharply with the figures in Section \ref{sec:singles} because much smaller eccentricities are sufficient for a run to be stopped, where in the previous cases the unstable configurations typically grew to $e_1 \geq 1$.    The maximum eccentricity, $e_{2,max}$, did not increase much over the initial value ($e_{2,o} \sim$0.05) for the stable cases.

\section{Conclusions}\label{sec:conc}
The stability of Earth-mass planets in the $\alpha$ Centauri AB system depends more strongly on the planet's initial eccentricity vector than was assumed in Paper I.  This is evident when we measure the maximum value of eccentricity and vary the initial longitude for planets orbiting either stellar component.  The binary orbit from \cite{Pourbaix2016} has shifted the dynamics slightly, where these changes largely reside within the chaotic regimes that are dependent on the starting semimajor axis $a_1$ and initial longitude of the planet.  

We determine numerically a relation for $\alpha$ Cen where the forced eccentricity can be estimated in either a prograde or retrograde direction relative to the orbital motion of the binary.  A piece-wise quadratic formula of the forced eccentricity $e_F$ as a function of $a_1$ is appropriate for most practical applications of prograde planets, where retrograde planets are well approximated using a linear function in $a_1$.

Stable planets with lifetimes of 1 Gyr occur at larger values of $a_1$ when they begin  at or near the corresponding forced eccentricity.  This results in a slight difference ($\sim$0.1 -- 0.3 AU) in the largest stable semimajor axis when considering prograde orbits around either star, but more substantial differences ($\sim$ 1 AU) occur for planets in retrograde orbits.  Retrograde planets can occupy stable orbits with $a_1$$\sim$5 AU for at least one billion years when $e_1 \approx e_F$.  

The forced eccentricity of circumbinary planets in $\alpha$ Cen is small; thus we simulated initially circular orbits in either a prograde or retrograde direction as in Paper I.  The differences in our circumbinary results on stability as a function of $a_1$ stem from the updated orbit (e.g., stellar masses) by \cite{Pourbaix2016}.  

The existence of a pair of planets alters the prospects of stability and system lifetime becomes more sensitive to the assumed initial eccentricity vector for planets on prograde circumstellar orbits.  When both planets begin near their forced eccentricities, the pair can survive up to much larger values of semimajor axis.  We performed a very narrow investigation of multiple planets in $\alpha$ Cen for this work, but present a more intensive study in \cite{Quarles2017b}.

\acknowledgments
{We thank the anonymous referee for providing helpful comments that improved the overall quality and clarity of the manuscript.}  The simulations presented here were performed using the OU Supercomputing Center for Education \& Research (OSCER) at the University of Oklahoma (OU).




\bibliographystyle{aasjournal}
\bibliography{bibliography}

\clearpage

\begin{deluxetable}{cc} \label{tab:bin}
\tablecolumns{2}
\tablewidth{0pc}
\tablecaption{Masses and Starting Orbital Elements of the Binary Stars}
\tablehead{\colhead{Element} & \colhead{Value}} 

\startdata
$a$($\prime\prime$) & 17.66 $\pm$ 0.026 \\
$a$(AU)$^*$ & 23.78 $\pm$ 0.04 \\
$i$($^\circ$) & 79.32 $\pm$ 0.044 \\
$\omega$($^\circ$) & 232.3 $\pm$ 0.11 \\
$\Omega$($^\circ$) & 204.75 $\pm$ 0.087 \\
$e$ & 0.524 $\pm$ 0.0011\\
$P$(yr) & 79.91 $\pm$ 0.013 \\
$M_{\rm A}$(M$_\odot$) & 1.133 $\pm$ 0.0050 \\
$M_{\rm B}$(M$_\odot$) & 0.972 $\pm$ 0.0045 \\
\enddata

\tablecomments{Orbital ephemeris assumed for the binary orbit taken from \citealt{Pourbaix2016}. The smallness of the uncertainties in the parameters illustrate the high accuracy of the orbital solution.  $^*$The semimajor axis has been derived from other relevant quantities via Kepler's 3$^{\rm rd}$ law.}

\end{deluxetable}

\begin{deluxetable}{ccccc}
\tablecolumns{5}
\tablewidth{0pc}
\tablecaption{Forced Eccentricity Coefficients for Prograde Circumstellar (S-Type) Orbits}
\tablehead{ & \colhead{Star} &\colhead{$C_1$} & \colhead{$C_2$} & \colhead{$C_3$}}
\startdata
$a_1<2$ AU & $\alpha$ Cen A & --0.007  & 0.044 &  --0.002 \\
& $\alpha$ Cen B & --0.009 & 0.047 & --0.003 \\
$a_1 \geq 2$ AU & $\alpha$ Cen A & --0.027  & 0.123 &  --0.080 \\
& $\alpha$ Cen B & --0.030 & 0.130 & --0.085 \\
\enddata
\tablecomments{Coefficients for the determining the forced eccentricity, $e_F = C_1a_1^2 + C_2a_1 + C_3$ (eq.~1), as a function of the starting semimajor axis in a {prograde} orbit around each stellar component.}\label{tab:coeff_pro}
\end{deluxetable}

\begin{deluxetable}{cc}
\tablecolumns{2}
\tablewidth{0pc}
\tablecaption{Forced Eccentricity Coefficients for Retrograde Circumstellar (S-Type) Orbits}
\tablehead{ \colhead{Star} &\colhead{$C_1$} }
\startdata
$\alpha$ Cen A & 1.399  \\
$\alpha$ Cen B & 1.465  \\
\enddata
\tablecomments{Coefficients for the determining the forced eccentricity, $e_F = C_1\frac{a_1}{a_{bin}}\frac{e_{bin}}{1 - e_{bin}^2}$ (eq.~2), as a function of the starting semimajor axis in a {retrograde} orbit around each stellar component.}\label{tab:coeff_ret}
\end{deluxetable}

\end{document}